\documentclass[doublecol]{epl2} 

\newcommand{\beq}{\begin{equation}}
\newcommand{\eeq}{\end{equation}}
\newcommand{\beqa}{\begin{eqnarray}}
\newcommand{\eeqa}{\end{eqnarray}}

\title{Nernst effect and diamagnetic response in a stripe model of \mbox{superconducting cuprates}}

\author{Ivar Martin\inst{1} \and C. Panagopoulos\inst{2,3} }

\institute{                    
  \inst{1} Theoretical Division, Los Alamos National Laboratory, Los Alamos,
New Mexico 87545, USA\\
  \inst{2} Department of Physics, University of Crete and FORTH, 71003
Heraklion, Greece\\
\inst{3} Division of Physics and Applied Physics, School of Physical and Mathematical Sciences, Nanyang Technological
University, 6373616 Singapore
}
\pacs{74.72.Kf}{Pseudogap regime}
\pacs{74.25.fg}{Electric and thermal conductivity}
\pacs{74.81.Fa}{Inhomogeneous superconductors and superconducting systems, including electronic inhomogeneities}

\abstract{
We examine the possibility that the experimentally observed enhancement of superconducting (SC) fluctuations above the SC transition temperature in the underdoped cuprates is caused by $stripes$ -- an intrinsic electronic inhomogeneity, common to hole-doped cuprates \cite{zaanen,PR,Mach,tranquada,berg}.
By evaluating the strengths of the diamagnetic response and the Nernst effect within a striped SC model, we find results that are qualitatively consistent with the experimental observations \cite{ong}. We make a prediction for anisotropic thermopower in detwinned samples that can be used to further test the proposed scenario.}

\begin{document}

\maketitle

One of the central puzzles of the high-temperature cuprate superconductors is the nature of the $pseudogap$ regime (PG) that extends from the parent Mott insulator to at least optimal doping  and up to several hundred degrees Kelvin  \cite{timusk}.  
Though it is unlikely that all of the vast PG region is directly related to superconductivity, there is compelling experimental evidence of anomalously strong SC fluctuations within PG  \cite{Corson,ong}.
In particular, in the lightly doped cuprates, the SC fluctuation regime can extend up to several times SC transition temperature $T_c$.

Two phenomena have particularly high sensitivity to the presence of SC, even local or fluctuating one: the diamagnetism and the Nernst effect.  The diamagnetic part of field-induced magnetization can reveal the presence of SC inclusions, as they expel magnetic field,  even when surrounded by completely non-SC (metallic or insulating) environment.  The Nernst effect is a thermal analog of the conventional Hall effect -- the transverse voltage generation in response to the heat current flow in a magnetic field. The Nernst effect in metals is typically very weak; however, in the presence of SC it is greatly enhanced due to the thermal drift of SC vortices, which induces voltage via the Josephson relationship.
In cuprates, the normal quasiparticle (qp) contribution may not be negligible as well \cite{taillefer}; however, since the qp and the vortex contributions have different temperature dependences, and often different signs, it is possible to disentangle the SC and qp contributions.

The strong positive correlation between the Nernst effect and the diamagnetic response makes a compelling case for persistence of SC fluctuations up to temperatures exceeding several times $T_c$ in the underdoped La  and Bi families of cuprates \cite{ong}.
The very broad SC fluctuation regime defies explanation within the Gaussian fluctuations approach \cite{gaussian1,gaussian2, gaussian3}, which prompted the conjecture that instead of translationally invariant dynamical fluctuations, the SC response above $T_c$ is governed by the quasi-static SC ``blobs" \cite{ong}.  These blobs can preserve SC -- locally -- up to temperatures $T_c^*$ significantly higher than the $T_c$, at which  the SC phase-locking occurs \cite{KT}.  Below $T_c^*$, the nucleation of SC regions is also expected to reduce in-plane resistivity.  Experimental evidence for this reduction below an intermediate temperature scale, sometimes referred to as $T_2^*$ (to distinguish from scale $T^*$ that marks onset PG), indeed exists \cite{loram}.

\begin{figure}
\onefigure[width = 3.2 in]{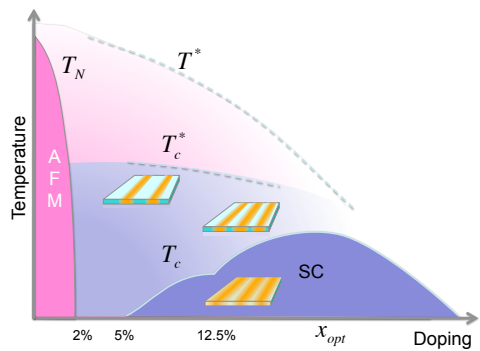}
\caption{(Color online) Schematic phase diagram.  Solid lines denote phase transitions into antiferromagnetic (AFM) and SC phases; dashed lines may either be phase transitions or crossovers, with $T^*$ denoting the onset of PG, as signaled by AFM fluctuations. The scale $T_c^*$ corresponds to the onset of local striped SC without global phase coherence.}
\label{FIG:PD}
\end{figure}

Strongly inhomogeneous SC may be caused by structural or chemical disorder. 
Alternatively, it can be induced by the $intrinsic$ spin/charge stripe inhomogeneity, which has been found across many classes of hole-doped cuprates \cite{ando,STM1,STM2, fujita}.  
That superconductivity follows the stripe pattern as well  (Fig. \ref{FIG:PD}), is indeed suggested by recent experiments  \cite{stripesSCnew1, stripesSCnew2}. The separation $\ell$ between stripes in the underdoped regime  is controlled by doping $x$.  For our estimates, we will assume that $\ell \approx (2x)^{-1}$ and will neglect peculiar features, such as the enhanced stripe stability in the vicinity of 1/8 doping. Temporal stripe fluctuations \cite{KFE} are not important as long as their timescale is slower than the characteristic timescale associated with SC, $\sim\hbar/T_c$. Similar to the SC blob scenario \cite{ong}, the global SC transition at $T_c$ is associated with the phase-locking of the SC order parameters among the stripes, while on individual stripes SC correlations can persist up to the local mean-field transition temperature $T_c^*$ \cite{MPK1,MPK2}.   That is, in the interval of  temperatures $T_c< T< T_c^*$, the stripes remain locally superconducting, while inter-stripe regions become normal.
At low doping (below $\sim$12\%), experimental evidence indicates that charge/spin stripes are well defined \cite{STM1,STM2} and thus SC is expected to be strongly inhomogeneous. 
At higher doping, the overlap between stripes becomes significant, leading to a nearly uniform SC below $T_c$.

\begin{figure}
\onefigure[width = 3.2 in]{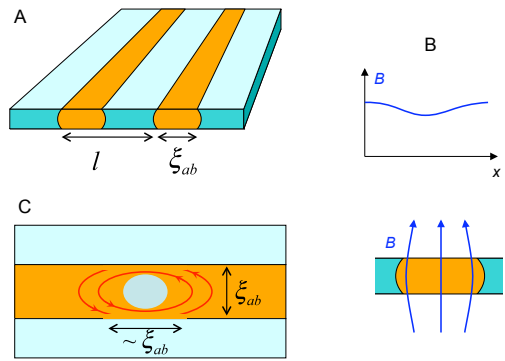}
\caption{(Color online) (A) Schematic of a striped SC below optimal doping and above $T_c$.  Dark (orange) regions are locally superconducting. (B) Weak expulsion of external magnetic field from the interior of the stripe due to the Meissner effect.  (C) Vortex penetration into the SC stripe, circulating (red) arrows denote diamagnetic screening currents. }
\label{FIG:str}
\end{figure}

Knowing that near optimal doping the striped SC should evolve into an almost uniform SC,  we can relate the SC parameters of the {\em stripe interior} with the {\em average} -- bulk -- parameters at optimal doping. We will assume that striped superconductor can be modeled as a series of SC wires with crossection $\xi_{ab} \times \xi_c$, with the respective SC coherence lengths taken to be the same as their measured (bulk) values at optimal doping. Such construction is consistent with having nearly uniform SC at optimal doping, since so defined in-plane width of the SC stripe $\xi_{ab} \sim 4 a$ is about the same as the known interstripe separation above 12\% hole doping. At lower doping, in our model the SC wires will be separated by non-superconducting regions, whose width increases with decreasing doping.
 In the direction along the stripes the SC correlation length is taken as $\xi_{ab}$ (Fig. \ref{FIG:str}A). 
 
\section{ Diamagnetism}
When SC stripes are exposed to magnetic field $B$, the diamagnetic surface currents act to partially expel the field (Fig. \ref{FIG:str}B).  The relevant penetration depth is $\lambda_{ab}(T)$ of the SC stripe interior (taken to be the same as the average penetration depth at optimal doping, which is  about $200$ nm  at $T = 0$ \cite{tallon}). Since the in-plane width of the stripe is much less than the penetration depth, $\xi_{ab} \ll \lambda_{ab}$, the fraction of the flux expelled from the interior of the stripe is $(\xi_{ab}/\lambda_{ab})^2$.  Averaging the diamagnetic response over SC and non-SC regions of space leads to an additional factor $\xi_{ab}/\ell$.  Therefore, the result for the SC stripe-induced diamagnetic contribution to susceptibility above $T_c$ is
\beq
\chi = \frac{\mu_0 M_{dia}}{B}\sim -\frac{2 \,\xi_{ab}^3(T) }{\lambda_{ab}^2(T) a}\,x,\label{eq:chi}
\eeq
where $\xi_{ab}(T)$ and $\lambda_{ab}(T)$ are functions of temperature, but not doping -- since they describe the SC properties of the stripe interior. 
The experimentally observed diamagnetic response in cuprates  above $T_c$ obeys the following pattern as a function of $B$, independent of particular material \cite{lulinew}:  $M \propto -B$ at small field, followed by a peak in $-M(B_p)$ at a temperature and doping dependent magnetic field $B_p(x,T)$, with the consequent disappearance of the diamagnetism at a higher field $B_h(T)$. Generically, within the striped SC scenario, when the coupling between the stripes can be neglected (for high enough $T$ and/or $B$), we expect the following simple scaling of diamagnetism: $M_{dia}(x_1,B,T)/M_{dia}(x_2,B,T) = x_1/x_2$. $B_h(T)$ would then correspond to the destruction of SC within the stripes, which we expect to be doping independent, as long as there are stripes. This indeed appears to be the case \cite{lulinew}: E. g., for temperatures below $T_c^{opt}$, $B_h$ is about 50 T for Bi-2201, 70 T for Bi-2212, and 60 T for La-214. These values remain approximately constant from optimal to underdoped samples, possibly enhanced near 1/8 doping, while  $T_c$ changes by factor 2-3. At low applied fields, the measured linear diamagnetic susceptibility within a given family of compounds increases with doping \cite{lulinew}. This behavior, as well as the magnitude of the effect, are qualitatively consistent with our estimate in Eq.~(\ref{eq:chi}).

\section{Nernst effect}
When in addition to magnetic field $B || \hat z$, an in-plane temperature gradient is applied, e. g., $\nabla T || \hat x$, a transverse electric field $E_y$ is generated.  The Nernst signal is defined as $N(B,T) = E_y/(\nabla_x T)$ in the absence of electrical current flow. In general, there are both normal (qp) and SC contributions to the Nernst effect.  The SC Nernst contribution has a natural interpretation in terms of thermally driven vortex motion  \cite{jja}. 
Within our model of the striped SC above $T_c$, the vortices are defined only while they tunnel across the SC regions. Since the vortex core size is about the same as the width of the SC stripe, $\xi_{ab}$, during the vortex tunneling nearly whole cross section of the SC stripe becomes normal (Fig. \ref{FIG:str}C). As the amplitude of the SC order parameter vanishes, the SC phase can wind by $\pm 2\pi$, leading, according to the Josephson relationship, to a voltage pulse between the ends of the SC stripe. This {\em phase-slip} dynamics has been proposed as the mechanism of electrical resistivity in thin superconducting wires  \cite{LAMH1,LAMH2}. In the case of a current-biased wire, the activation barrier for the phase slips, which generate positive voltage pulses, is lower than for the anti-phase slips.   Upon time averaging, this leads to positive DC resistivity.

In the Nernst measurement, there is no applied current; instead, the symmetry between vortex and anti-vortex tunneling across the SC stripes is broken by the external magnetic field. The vortices with magnetic moment $m$ aligned with the external magnetic field have lower energy than anti-vortices, which carry moment $-m$.  Thus, the activation barrier for a vortex/antivortex crossing a thin SC wire is $U_{v/av} = U_0 \mp m B$. The SC condensation energy within the region quenched by the vortex core is $U_0 \approx  (1/\mu_0)B_c^2 \xi_{ab}^2 \xi_c$, with the thermodynamic critical field $B_c \approx \phi_0/(\lambda_{ab}\xi_{ab})$ and $\phi_0 = h/(2e)$ the flux quantum. The thermodynamic critical field is related to the second critical field  \cite{degennes}, $B_c \approx B_{c2} \xi_{av}/\lambda_{ab} $.  The magnitude of the vortex/antivortex magnetic moment $m$ can be estimated by noticing that the magnetization density near the vortex core scales as $(1/\mu_0)\phi_0/\lambda_{ab}^{2}$ and thus the total magnetic moment of a vortex confined to a SC wire of width $\xi_{ab}$ and thickness $\xi_c$ is $m \sim (1/\mu_0)\phi_0 \xi_{ab}^2 \xi_c/\lambda_{ab}^{2}$. Therefore, for the tunneling energy barrier we obtain $U_{v/av} = U_0(1 \pm B/ {\cal B})$, with ${\cal B} \sim \phi_0/\xi_{ab}^2 \approx B_{c2}$, the intrastripe second critical field (approximately equal to the second critical field at optimal doping).  

Let us first assume that the stripes are perpendicular to the temperature gradient. Then, the temperature differential across the SC stripe is $\Delta T \approx |\nabla T| \xi_{ab}$.  The activated tunneling rate for either vortices or antivortices is $\Gamma = \Gamma_0 e^{-U/T} $, where the prefactor can be estimated  \cite{LAMH2} as $\Gamma_0 \sim T_c^*$ at temperatures below but not very near $T_c^*$. The temperature difference across the stripe causes asymmetry in the vortex flow, with the vortex/antivortex tunneling rate from hot to cold exceeding the rate of going in the opposite direction (``entropic forcing").
For the tunneling approach to apply, the barrier height $U$ has to be much larger than temperature. Combining the vortex and antivortex tunneling contributions over the elementary stripe segments of length $\xi_{ab}$, and using the Josephson relationship, for the average electric field along the stripe we find,
\beq
N = \frac{E_y} {\nabla_x T}\sim \Gamma_0\phi_0 e^{-\frac{U_0}{T}}\frac{U_0}{T^2}\left({\rm sh} \frac{U_0 B}{T B_{c2}} - \frac{B}{B_{c2}}{\rm ch} \frac{U_0 B}{T B_{c2}}\right).\label{eq:Ey}
\eeq
For small magnetic fields, $B < B_{c2} T/U_0$, the dependence is linear
\beqa
\nu = \frac{E_y}{B\nabla_x T} \sim\Gamma_0\phi_0\,\frac{U_0^2}{T^3 B_{c2}}\exp{-\frac{U_0}{T}}\\
\sim \frac{k_B}{e B_{c2}}\left(\frac{U_0}{T}\right)^2 \exp{-\frac{U_0}{T}}.
\eeqa
For a typical value of $B_{c2}\sim 100$ T, we find that the first factor in the product is about $10^{-6} $ V/(K T), while the rest of the expression is very strongly dependent on the value of $U_0/T$. For instance, if $U_0/T = 5$, we find $\nu \sim 10^{-7}$ V/(KT).

\begin{figure}
\onefigure[width = 3.2 in]{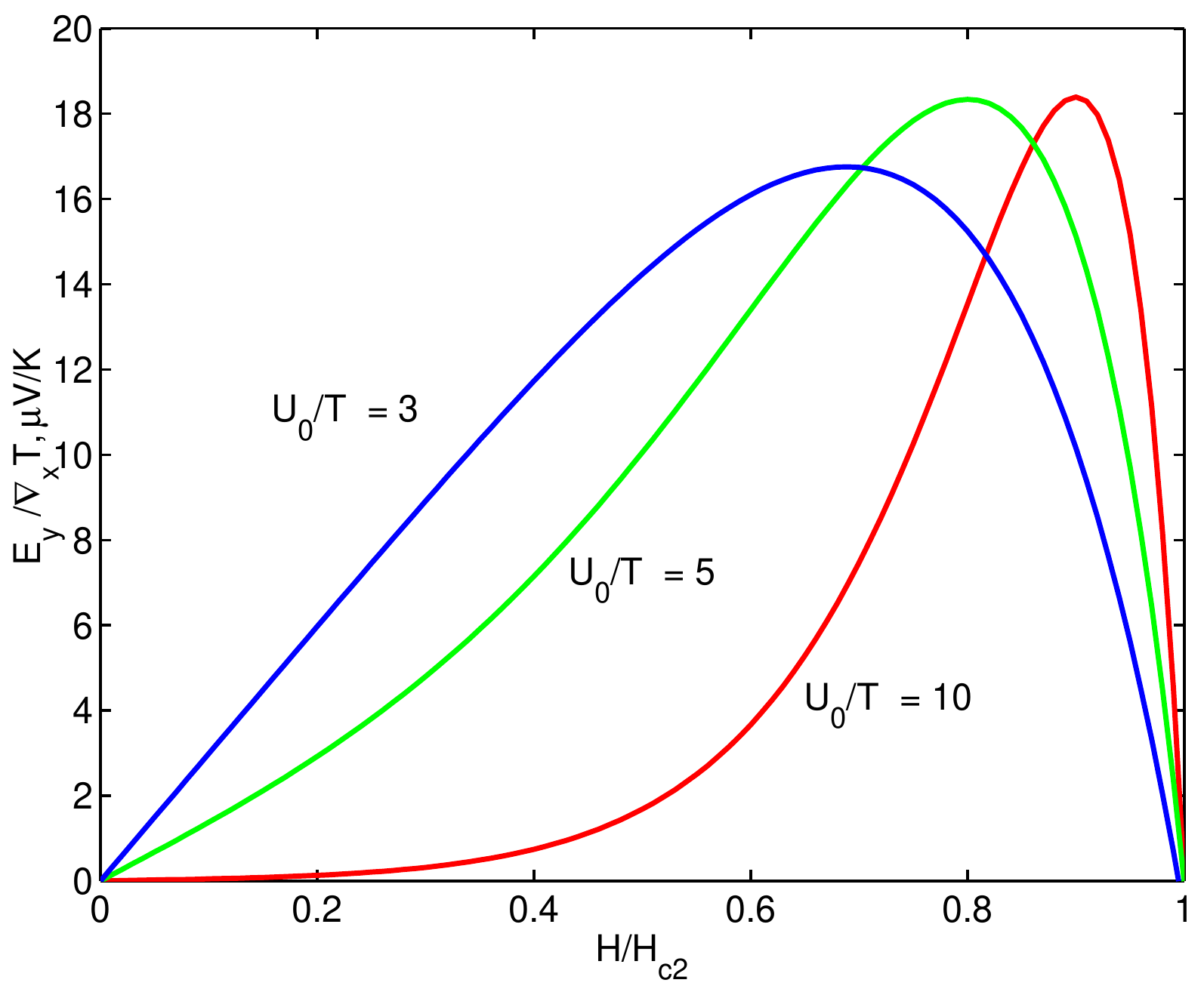}
\caption{(Color online) Theoretical prediction of the Nernst signal based on the expression of Eq.~(\ref{eq:Ey}) for several values of the ratio of the zero-field vortex tunneling barrier to temperature, $U_0/T$.}
\label{FIG:Nernst}
\end{figure}

For $U_0/T \gg 1$, as $B$ exceeds $B_{c2} T/U_0$, from Eq.~(\ref{eq:Ey}), the Nernst signal should increase non-linearly, reaching the nearly $U_0$-independent value $N_{max} \sim 0.2 k_B/e \sim 20\ \mu$V/K below $B_{c2}$ and then drop rapidly to zero at $B_{c2}$.
Several theoretical curves for the Nernst signal at different temperatures are presented in Fig.~\ref{FIG:Nernst}.

For  comparison with the experiment, we focus on the Bi and La families of cuprates  \cite{ong}. 
The main features of the measured Nernst behavior are closely analogous to  the  diamagnetic response described above. Namely, $N(B)$ has a pronounced dome-like shape, with the signal concentrated within an interval of magnetic fields that does not significantly depend on the doping.  This is naturally consistent with the striped SC scenario, according to which the doping controls separation between the stripes but not their internal SC properties. Moreover, at low temperatures,  a non-linear dependence $N(B)$ is observed, as expected from  Eq.~(\ref{eq:Ey}). 
While we cannot claim quantitative agreement between Eq.~(\ref{eq:Ey}) and the experimental results, we can roughly estimate the theoretically expected magnitude of the Nernst signal. The main uncertainty is the value of $U_0$ since it appears in the exponent. Given the known optimal-doping properties of the cuprates, we estimate $U_0(T = 0) < 500$ K.  Experimentally, the form of the Nernst signal as a function of $B$ and its magnitude would be consistent with having $U_0/T \sim 1\ldots 3$ in Eq.~(\ref{eq:Ey}). These values of $U_0/T$ are at the border of validity of the tunneling approximation that we used here; this may indicate the ``optimal tuning" of superconductivity in cuprates, i.e., that  stripes 
 achieve an optimal compromise between the paring strength and superfluid stiffness.

Above we assumed that the stripes are ordered and continuous, running perpendicular to the direction of the heat flow, and connecting the sides of the sample, between which  the transverse voltage is measured.
Let us now consider more general situation when the 
angle $\theta$ between  the stripes and the direction of the applied thermal gradient ($\hat x$) is arbitrary. The temperature differential across SC stripe that drives the vortex current will be reduced by the factor $\sin\theta$. However, the length of stripe which connects the sides of the sample and is available for vortex tunneling will increase in the same proportion,  $\Delta \ell = \Delta y/\sin\theta$. We therefore conclude that the transverse electric field $E_y$ is independent of the stripe orientation, {\em as long as stripes connect the two sides of the sample}. Naturally, if the stripes run in the direction of the heat current, the Nernst signal will vanish since none of the stripes will connect the sides of the sample.

A direct consequence of the striped SC scenario for the Nernst effect is that in the samples where stripes run predominantly in one direction, there should also be anisotropic SC contribution to thermopower that is odd in magnetic field. That is because the electric field generated  by the phase slips is parallel to the stripes; therefore, if stripes are not orthogonal to $\nabla_x T$, in addition to the transverse field $E_y$ there will be a  $longitudinal$ field 
\beq
E_x = E_y \cot\theta.
\eeq
Such strongly anisotropic response implies that the results of the experimental measurement may depend on the sample shape and the positioning of the contacts. Qualitatively, however, thermopower $E_x$ that is odd in $B$ and $\theta - \pi/2$, and comparable in magnitude to the Nernst electric field $E_y$  is expected (by symmetry, $E_x = 0$ at $\theta = 0$). This prediction should be possible to test, e.g., in detwinned YBCO samples.

We now briefly mention other observable implications of the striped SC scenario. The inhomogeneous superfluid density due to stripes will have signatures in the AC conductivity and in the average penetration depth. Below $T_c$, at low doping, the superfluid stiffness is expected to be larger in the direction along the stripes (scaling as $x$), and smaller perpendicular to the stripes (scaling as $e^{-1/x}$). This anisotropy will lead to redistribution of the spectral weight of the real part of conductivity between the SC delta-function peak at zero frequency and the finite frequencies  \cite{barabash}, when measured perpendicular to the stripes. Such spectral weight transfer is well known to occur in underdoped cuprates \cite{oren2}. Through $T_c$, the global phase coherence is lost; however, the in-plane AC conductivity is expected to change gradually, with the SC signatures disappearing when the excitation frequency falls below the characteristic ``phase slip" frequency of SC fluctuations  \cite{koshelev}. The influence of this additional time scale on the optical conductivity measurements may account for the discrepancy between this and other means (e.g. Nernst and diamagnetism) of identifying the ranges of superconducting fluctuations.

So  far we did not discuss the role of stripes in promoting or suppressing superconductivity, beyond providing a template to which it has to adjust. That inhomogeneity may be beneficial for superconductivity in cuprates follows from several conspicuous correlations: SC coherence length similar to the interstripe separation near optimal doping, and 
large overlap between the regions of the SC and striped regions of the phase diagram.
In fact, the highest onset temperature for the Nernst signal and diamagnetism in LSCO is at 12\%  \cite{ong}, precisely where the stripes are the most robust due to commensuration effects with the lattice.
Further insight into the question of stripe relevance to SC could be obtained by correlating the high-temperature onset of incommensurate neutron and X-ray scattering with the signatures of SC fluctuations \cite{tranqnew}.

To conclude, we considered a qualitative model of striped superconductivity in an attempt to understand the extraordinarily broad superconducting fluctuation region above $T_c$ in underdoped cuprates.  Even though our model likely exaggerates the role of stripes it provides a complementary framework to the more common homogeneous approaches, and thus may help to develop comprehensive understanding of these materials in the future.

\acknowledgments

We acknowledge useful discussions with C. Kallin, D. van der Marel, and  D. Podolsky, and thank Aspen Center for Physics, where this work was initiated, for hospitality. The work of IM
was carried out under the auspices of the National Nuclear Security
Administration of the U.S. Department of Energy at Los Alamos National
Laboratory under Contract No. DE-AC52-06NA25396 and supported by the LANL/LDRD
Program. CP acknowledges financial support from MEXT-CT-2006-039047, EURYI and the National Research Foundation, Singapore.

\end{document}